\documentstyle[prb,aps,multicol,epsf]{revtex}
\twocolumn
\begin{document}
\draft
\title{Nonadiabatic noncyclic geometric phase of a spin-$\frac{1}{2}$
particle subject to an arbitrary magnetic field}
\author{Shi-Liang Zhu and  Z. D. Wang$^*$ }
\address{Department of Physics, University of Hong Kong,
Pokfulam Road,
Hong Kong, P. R. China\\}
\author{Yong-Dong Zhang}
\address{ Department of Automatic Control Engineering,
South China University of Technology, Guangzhou, 510641, P. R. China}
\address{\mbox{}}
%\date{\today}
\address{\parbox{14cm}{\rm \mbox{}\mbox{}
We derive a formula of the nonadiabatic noncyclic Pancharatnam phase 
for a quantum spin-$\frac{1}{2}$ particle
subject to an arbitrary magnetic field. 
The formula is applied to three specific kinds of magneic fields.
(i) For an orientated magnetic field, the Pancharatnam phase
is derived exactly.
(ii) For a rotating magnetic field,
the evolution equation is solved analytically. The
Aharonov-Anandan phase is obtained exactly and  the Pancharatnam
phase is computed numerically. (iii) We propose
a kind of topological transition in one-dimensional
mesoscopic ring subject to an in-plane magnetic field,
and then address the nonadiabatic noncyclic effect on
this phenomenon.
}}
\address{\mbox{}}
\address{\parbox{14cm}{\rm PACS numbers: 03.65.Bz}}
\maketitle

%\newpage
%\narrowtext

\section{Introduction}

Berry's phase~\cite{Berry} and its generalization, the Aharonov-Anandan(AA)
phase~\cite{Aharonov}, have attracted considerable attention in 
recent years~\cite{Shapere}. It was discovered by Berry that a geometric phase
$\gamma_n(C)=i\oint_C\langle n(\stackrel{\rightarrow}{R})
|\nabla_{\stackrel{\rightarrow}{R}}|n(\stackrel{\rightarrow}{R})\rangle \cdot
d \stackrel{\rightarrow}{R}$, in addition to the usual dynamic phase,
$-\frac {1}{\hbar}\int_0^{\tau}E_n(\stackrel{\rightarrow}{R})dt$,
is accumulated on the wavefunction of a quantum system, provided that the 
Hamiltonian is cyclic and adiabatic. This adiabatic geometric phase
has found many applications in
physics, particularly in mesoscopic
systems where the quantum interference is
important. Loss
{\sl et al} found that the persistent currents can be induced by 
the adiabatic Berry phase in a {\sl closed} mesoscopic ring embedded in
a static inhomeneous magnetic field~\cite{Loss}. 
Zhu {\sl et al} proposed a novel
experiment to test AA phase in a textured mesoscopic {\sl open}
ring subject to a crown-like magnetic field~\cite{Zhu}. An 
interesting kind of topological transition induced by the interference of 
the adiabatic Berry phase
was proposed in Ref.~\cite{Lyanda-Geller}. Moreover,
the geometric phase can be generalized to even noncyclic
evolution~\cite{Samuel,Li,Jordan}, and a very recent experiment to test
the noncyclic evolution is reported by Wagh {\sl et al}~\cite{Wagh}.

    While dealing with the interference of light, Pancharatnam came
up with a brilliant idea regarding a general phase of
the evolution for a polarized light~\cite{Pancharatnam},
which was then generalized
to an arbitrary quantum evaluation~\cite{Samuel,Berry2}. When
a system evolves 
from an initial state $|\psi (0)\rangle$
to a final state $|\psi (t)\rangle=\hat{U}(t)|\psi (0)\rangle$ 
with $\hat{U}(t)$ a unitary evolutation operator and 
$\langle\psi (0)|\hat{U}(t)|\psi (0)\rangle\not=0$, we refer 
$\gamma_t$ as the phase of $|\psi(t)\rangle$ relative to $|\psi(0)\rangle$
once we have 
\begin{equation}
\label{Pancharatnam}
\langle\psi(0)|\hat{U}(t)|\psi(0)\rangle=e^{i\gamma_t}|
\langle\psi(0)|\hat{U}(t)|\psi(0)\rangle|.
\end{equation}
For an arbitrary quantum evolution, the geometric 
Pancharatnam phase can be defined as $\gamma_p=\gamma_t-\gamma_d$, where
$\gamma_d
=-\frac {1}{\hbar}\int_0^t \langle\psi(t)|\hat{H}
|\psi(t)\rangle dt$ is the dynamical phase with
$\hat{H}$ as the Hamiltonian of the system.

   Consider a quantum system whose normalized state vector $|\psi(t)\rangle$
evolves according to the Schr\"{o}dinger equation
$i\hbar\frac{d}{dt}|\psi(t)\rangle=\hat{H}
(t)|\psi(t)\rangle$ . Let us define a new state vector 
$|\phi(t)\rangle$ which differs from $|\psi(t)\rangle$ only
in that its dynamical phase factor has been removed.
The Pancharatnam phase
difference between any two nonorthogonal elements of $\aleph$ can be
obtained by the following geodesic rule: If one writes 
$\langle\phi_1|\phi_2\rangle=\rho exp(i\gamma)$, $\rho>0$, 
the phase $\gamma$ is given by the line integral of
$A_s$ along any geodesic lift from $|\phi_1\rangle$
to $|\phi_2\rangle$~\cite{Samuel}, where
$A_s=Im\langle\phi(s)|d/ds|\phi(s)\rangle$ with
$s$ as a parameter.
Using this rule, we are able to calculate the nonadiabatic
noncyclic Pancharatnam phase accumulated in the evolution of a
spin-$\frac{1}{2}$ particle
subject to an arbitrary magnetic field;
It is worth noting that the Pancharatnam phase
has physical reality only when the rotated part of the
wave function is somehow made to interfere with another part
that was not rotated.
The formulas to be derived can be used for all two-level systems
because any two-level system can be mapped into 
a system of the spin-$\frac{1}{2}$ in a
specific magnetic field~\cite{Feynman}.

  On the other hand, with the advancement
of nanotechnology, it is
possible to fabricate mesoscopic rings of size within the phase coherence 
length so that the phase memory is retained by electrons throughout the 
whole system. In such systems, the electronic 
quantum transport is significantly affected by
the geometric phase which may not be cyclic or adiabatic,
However,
most theoretical studies of the geometric phase
in mesoscopic systems have so far
been limited to the cases of adiabatic or cyclic electronic transport.
Therefore, it is quite useful and
interesting to investigate theoretically the noncyclic nonadiabatic 
geometric phase and its effect on the electronic transport
in mesoscopic systems.
Motivated by this, we study
the noncyclic nonadiabatic Pancharatnam phase of an electron and
discuss the related quantum inference   
in a mesoscopic ring connected to current leads subject to 
a magnetic field.

The paper is organized as follows. In Sec. II, we derive
a formula of the noncyclic 
nonadiabatic geometric Pancharatnam phase for a 
quantum particle of spin-$\frac {1}{2}$  subject to an arbitrary 
magnetic field. In Sec. III, the formula is
applied to the three systems subject to, respectively,
three specific magnetic fields.
For an orientated magnetic field,
the Pancharatnam phase is derived exactly.
For a rotating magnetic field, 
the evolution equation is solved analytically, and the geometric phase is
computed numerically.
In particular, a striking topological
transition in a mesoscopic ring subject to an
in-plane magnetic field is addressed.
The paper ends with a brief
summary.

\section{General formula}

The Hamiltonian for a system of
spin-$\frac {1}{2}$ particle subject to
an arbitrary magnetic field ${\bf B}(t)$ is given by
\begin{equation}
\label{Hamiltonian}
\hat{H}(t)=-\frac{\mu}{2}{\bf B}(t)\cdot
\stackrel{\rightarrow}{\sigma},
\end{equation}
where $\mu$ is the Bohr magneton, and
$\stackrel{\rightarrow}{\sigma}=(\sigma_x,\ \sigma_y,\ \sigma_z)$ with
$\sigma_{x,y,z}$ as Pauli matrices.
The space of states of this system is the projective space $CP^{(1)}$, which
is diffeomorphic to the unit sphere $S^2$
($CP^{(1)}\simeq S^3/U(1)\simeq S^2$). The point in $S^2$ associated
with an arbitrary state $|\psi\rangle$ of the system is 
${\bf n}=\langle\psi|\stackrel{\rightarrow}{\sigma}
|\psi\rangle$. Reciprocally, for a given vector 
${\ bf n}\in S^2$,
parameterized in a North chart by
$$
{\bf n}=(n_1,n_2,n_3)=
(sin\theta cos\varphi, 
sin\theta sin\varphi,cos\theta),
$$
we can associate this vector with the spin state
$$
|\psi\rangle=
\left (
\begin{array}{l}
e^{-i\varphi/2}cos(\theta/2)\\
e^{i\varphi/2}sin(\theta/2)
\end{array} 
\right )_\sigma,
$$
where subscript $\sigma$ denotes the spin space.
The Schr$\ddot{o}$dinger equation for the state
$|\psi(t)\rangle$,  
$\frac{d}{dt}|\psi(t)\rangle=-\frac{i}{\hbar}\hat{H}
(t)|\psi(t)\rangle$, 
can be expressed in the following form for the vector
${\bf n}(t)$: $\frac {d{\bf n}(t)}{dt}=
-\frac{\mu}{\hbar} {\bf B}(t)\times
{\bf n}(t)$~\cite{Wagh1}. This equation can be rewritten in
a matrix form as
\begin{equation}
\label{evolution}
\frac {d{\bf n}^{T}(t)}{dt}=\hat{B}_M (t){\bf n}^{T}(t),
\end{equation}
with 
$$
\hat{B}_M(t)=\frac{1}{\hbar}\left (
\begin{array}{lcr}
0 & \mu B_3(t) & -\mu B_2(t)\\
-\mu B_3(t) & 0 & \mu B_1(t)\\
\mu B_2(t) & -\mu B_1(t) &  0\\
\end{array}
\right )
$$
for ${\bf B}(t)=
(\begin{array}{lcr} B_1(t),& B_2(t),&B_3(t)\end{array})$,
where $T$ represents the transposition of matrix.

The evolution from an initial state ${\bf n}(0)$
to a final state ${\bf n}(t)$
corresponds to a curve on the sphere $S^2$.
This field-depend curve may be very complicated. 
A cyclic evolution of the state is represented by
a closed curve on the sphere, that
is, ${\bf n}(\tau)={\bf n}(0)$
with $\tau$ as a period of a cycle. 
Whether the evolution is cyclic or not is dependent on both
the magnetic field and the initial state.
The evolution of the spin-$\frac{1}{2}$
system is noncyclic in general although it is cyclic in 
some special cases, which we will discuss later on.
The general curve ${\bf n}(t)$   
can hardly be solved analytically, even though 
${\bf n}(t)$
may be exactly determined in some special conditions.
The solution of Eq.(\ref{evolution})
may be written formally as a $\hat{T}$-exponential:
${\bf n}^T(t)
=\hat{T}exp(\hat{Q}(t)){\bf n}^T(0)$ with 
$\hat{T}$ as the time-ordering operator and
$\hat{Q}(t)=\int_0^t \hat{B}_M(t')dt'$.
We can ignore the $\hat{T}$-operator if $\hat{B}_M(t)$ at different times commute. 
Once we find an operator $\hat{S}(t)$ to diagonalize $\hat{Q}(t)$ in the base:
$\hat{I}(t)=\hat{S}^{-1}(t)\hat{Q}(t)\hat{S}(t)
=diag(\lambda_1(t),\lambda_2(t),\lambda_3(t))$, 
we have the exact
solution:
\begin{equation}
\label{exact-sol}
{\bf n}^T(t)
=\hat{S}(t)e^{\hat{I}(t)}\hat{S}^{-1}(t) {\bf n}^T(0). 
\end{equation}

For a general initial state
$$
|\phi(t_i)\rangle
=\left (\begin{array}{l}
e^{\frac {-i\varphi_i}{2}}cos\frac {\theta_i}{2}\\
e^{\frac {i\varphi_i}{2}}sin\frac {\theta_i}{2}
\end{array} \right )_\sigma,
$$
the state at the instant $t$ is 
$$
|\phi(t)\rangle
=\left (\begin{array}{l}
e^{\frac{-i\varphi_{(t)}}{2}}cos\frac {\theta(t)}{2}\\
e^{\frac{i\varphi_{(t)}}{2}}sin\frac {\theta(t)}{2}
\end{array} \right)_{\sigma}.
$$
A unique curve ${\bf n}(t)$ on the unit sphere
$S^2$ is determined by the evolution $|\phi(t)\rangle$
with the initial point $A$ of coordinates 
${\bf n}(t_i)=
(sin\theta_i cos\varphi_i, 
sin\theta_i sin\varphi_i,cos\theta_i)$
and the final point $P$ of coordinates
${\bf n}(t_f)
=(sin\theta_f cos\varphi_f, 
sin\theta_f sin\varphi_f,cos\theta_f)$.
Then, 
$|\phi(t)\rangle=\hat{U}(t,0)|\phi(0)\rangle$ with 
$\hat{U}(t,0)=\hat{T}exp(-\frac{i}{\hbar}\int_0^t\hat{H}(t')dt')$
as the unitary evolution operator
which gives a curve $\overbrace{AHP}$
on the unit sphere $S^2$.
If $\langle\phi(0)|\hat{U}(t_f,0)|\phi(0)\rangle$ is not zero,
the Pancharatnam phase $\gamma_p(t_f)$ is defined by
\begin{equation}
\label{P-phase0}
\langle\phi(0)|\hat{U}(t_f,0)|\phi(0)\rangle
=e^{i\gamma_p(t_f)}|\langle\phi(0)|\hat{U}(t_f,0)|\phi(0)\rangle|.
\end{equation}
Clearly, $\gamma_p(t_f)$
recovers the AA phase $\gamma_{AA}$ if ${\bf n}(t_f)=
{\bf n}(0)$ for $t_f=\tau>0$~\cite{Aharonov}.
For a noncyclic evolution,
we can introduce a specific 
unitary operator $\hat{U}_c(\tau,t_f)$ which makes 
${\bf n}(\tau)={\bf n}(0)$ 
after the evolution
$|\phi(\tau)\rangle=\hat{U}_c(\tau,t_f)|\phi(t_f)\rangle$, and thus we
have 
\begin{eqnarray}
\langle\phi(0)|\hat{U}(t_f,0)|\phi(0)\rangle 
& = & \langle\phi(0)|\hat{U}^{+}_c(\tau,t_f)
\hat{U}_c(\tau,t_f)\hat{U}(t_f,0)|\phi(0)\rangle \nonumber \\
\label{P-phase1}
& = & \langle\phi(0)|\hat{U}_c^{+}(\tau,t_f)|\phi(\tau)\rangle \nonumber\\
& = & \langle\phi(0)|
\hat{U}_c^{+}(\tau,t_f)|\phi(0)\rangle e^{i\gamma_{AA}(\tau)}. 
\end{eqnarray} 
If $\langle\phi(0)|\hat{U}_c^{+}(\tau,t_f)|\phi(0)\rangle$ is
real and positive,
it is clear from Eqs.(\ref{P-phase0}) and (\ref{P-phase1}) that
the Pancharatnam phase for the noncyclic evolution is given by
the AA phase of the specific cyclic evolution $C$
determined by the operator $\hat{U}_c(\tau,t_f) \hat{U}(t_f,0)$, {\sl i.e.},
\begin{equation}
\label{P-AA}
\gamma_p(t_f)=\gamma_{AA}(\tau).
\end{equation}

We now consider a special evolution operator $\hat{U}_g(\tau,t_f)$
which makes
the state pass from $P$ to $A$ along the shortest path
$\overbrace{PSA}$ ({\sl i.e.}, the geodesic curve)
in the unit sphere of $S^2$.
Then, $\overbrace{AHP}$ and $\overbrace{PSA}$ forms a closed
curve $C$ on the surface $S^2$. The geometric phase
for this cycle is determined from the surface area
$S_C$ closed by the curve $C$~\cite{Li2}, {\sl i.e.},
$$
\gamma_{AA}(\tau)=-\frac{1}{2}S_C
=-\frac{1}{2}\int_{Surface}
{\bf n}\cdot d{\bf S}.
$$
The surface integral can be transformed into a
line integral by introducing a vector
field ${\bf A}_s$ such that $\nabla\times
{\bf A}_s=-{\bf n}/2$
on the surface. It can be found that the vector potential 
${\bf A}_s({\bf n})
=(\frac{n_2}{2(1+n_3)},\frac{-n_1}{2(1+n_3)},0)$
describing
the field of a monopole  
satisfies the requirement~\cite{Wu}.
Therefore, we have
\begin{eqnarray}
\gamma_{AA}(\tau) &=& \oint_C
{\bf A}_s\cdot d{\bf n}\nonumber\\
\label{AA}
&=&-\frac{1}{2}\int_0^{tf} \frac
{n_1\stackrel{\cdot}{n}_2 - n_2\stackrel{\cdot}{n}_1}
{1+n_3}dt
+\int_{{\tiny \overbrace{PSA}}}{\bf A}_s\cdot d{\bf n},
\end{eqnarray}
where the dot denotes the time derivative, and the second line integral
is performed along the shorter geodesic curve from $P$ to $A$.
The equation to describe the geodesic curve through point $P$ to $A$ can be expressed as
\begin{equation}
\label{geodesic}
tg\theta=\frac {-\kappa}{\eta cos\varphi+\zeta sin\varphi},
\end{equation}
with
$$ 
\begin{array}{l}
\eta=n_2(t_i)n_3(t_f)-n_3(t_i)n_2(t_f),\\
\zeta=-n_1(t_i)n_3(t_f)+n_3(t_i)n_1(t_f),\\
\kappa=n_1(t_i)n_2(t_f)-n_2(t_i)n_1(t_f).
\end{array}
$$
Substituting Eq.~(\ref{geodesic})
into  
$\int_{{\tiny\overbrace{PSA}}}{\bf A}_s
\cdot d{\bf n}$,
we obtain
\begin{equation}
\label{PA}
\int_{{\tiny \overbrace{PSA}}}{\bf A}_s \cdot d{\bf n}=
arctg\frac {sin(\varphi_f-\varphi_i)}
{ctg\frac {\theta_f}{2} ctg\frac {\theta_i}{2}+cos(\varphi_f-\varphi_i)}.
\end{equation}
The evolution curve determined from the
above operator $\hat{U}_g(\tau,t_f)$ is the geodesic curve, which
indeed ensures
$\langle\phi(0)|\hat{U}_g^{+}(\tau,t_f)|\phi(0)\rangle$ to
be real and positive~\cite{Samuel,Jordan}.
Therefore, we have, from Eqs.~(\ref{P-AA}),~(\ref{AA}) and ~(\ref{PA}),
\begin{eqnarray}
\gamma_p(t_f)= &-& \frac{1}{2}\int_0^{tf} \frac
{n_1\stackrel{\cdot}{n}_2 - n_2\stackrel{\cdot}{n}_1}
{1+n_3}dt\nonumber\\
\label{phase}
&+& arctg\frac {sin(\varphi_f-\varphi_i)}
{ctg\frac {\theta_f}{2} ctg\frac {\theta_i}{2}+cos(\varphi_f-\varphi_i)}.
\end{eqnarray}
Equation (\ref{phase}) is a central result of this paper, which provides
a very useful formula for computing the
noncyclic nonadiabatic geometric phase for
any two-level system.
We emphasize that Eq.(\ref{phase}) can be used to any evolution of a 
spin-$\frac{1}{2}$ particle subject to
an arbitrary magnetic field ${\bf B}(t)$.

\section{Applications to three specific systems}

We now apply Eq.(\ref{phase}) to systems subject to
an orientated magnetic field, a rotating magnetic field, and a rotating plus
a constant magnetic field.

\subsection{An orientated magnetic field}

The simplest system is that a spin-$\frac{1}{2}$
particle is subject to an orientated
magnetic field, which can be written as 
${\bf B}=(0,0,B_3)$.
The Pancharatnam phase for this system can be 
obtained 
straightforwardly even for time-dependent $B_3$
because 
the magnetic
matrix $\hat{B}_M(t)$ at different times commute.
One can find that
$$
\hat{S}(t)e^{\hat{I}(t)}\hat{S}^{-1}(t)
=\left (
\begin{array}{lcr}
cos\varphi_t & -sin\varphi_t & 0\\
sin\varphi_t & cos\varphi_t & 0\\
0 & 0 & 1
\end{array}  \right )
$$
where $\varphi_t=-\frac {2\mu }{\hbar}\int_0^{t} B_3(t')dt'$.
Thus, for the initial state
${\bf n}(0)=
(sin\theta_i cos\varphi_i, 
sin\theta_i sin\varphi_i,cos\theta_i),$
we have
${\bf n}(t)=
(sin\theta_i cos(\varphi_i+\varphi_t), 
sin\theta_i sin(\varphi_i+\varphi_t),cos\theta_i)$
from Eq.(\ref{exact-sol}).
Therefore, 
it is straightforward from Eq.(\ref{phase}) to find
\begin{equation}
\label{phase-B3}
\gamma_p(t)=-\frac {\varphi_t}{2}(1-cos\theta_i)
+arctg\frac{sin\varphi_t}{ctg^2\frac{\theta_i}{2}+cos\varphi_t}.
\end{equation}
We can rewrite Eq.(\ref{phase-B3}) as
$$
tg[\gamma_p(t)-\frac{\varphi_t}{2}cos\theta_i]
=-tg\frac{\varphi_t}{2}cos\theta_i,
$$
which recovers the result for
the constant magnetic field $B_3$ reported in
Ref.\cite{Wagh}.
This
noncyclic geometric phase was
indeed detected
in a well-performed polarized neutron interferometric experiment.

\subsection{A rotating magnetic field}

Consider a spin-$\frac{1}{2}$ quantum particle in a rotating magnetic field.
The Hamiltonian of the system is Eq.(\ref{Hamiltonian})
with the magnetic field given by
\begin{equation}
\label{crown-shaped}
{\bf B}=(B_0cos\omega t, B_0sin\omega t, B_1),
\end{equation}
where $B_0$ and $B_1$ are constants.

The {\sl adiabatic and cyclic} Berry phase
for this system has been found to be 
$\mp \frac{1}{2}\Omega_C$ with $\Omega_C=2\pi(1-cos\alpha)$ as
the solid angle that $C$ subtends to the center of the unit sphere
\cite{Berry},
where $\alpha=arctg(B_0/B_1)$ is the fixed tilt angle.
A general evolution follows a nonadiabatic, and
even a noncyclic
one. The magnetic matrix $\hat{B}_M$ can now be expressed as
$$
\hat{B}_M(t)
=\left (
\begin{array}{lcr}
0 & \omega_1 & -\omega_0 sin\omega t\\
\omega_1 & 0 & \omega_0 cos\omega t\\
\omega_0 sin\omega t & -\omega_0 cos\omega t & 0
\end{array}  \right ),
$$
with $\omega_i=\mu B_i/\hbar$.
Though the matrices $\hat{B}_M(t)$ at different times do not commute, we can
still solve this problem exactly.
Let us introduce a new vector
\begin{equation}
\label{new-vector}
{\bf u}^T(t)
=\left (
\begin{array}{lcr}
cos\omega t & sin\omega t & 0\\
-sin\omega t & cos\omega t & 0\\
0 & 0 & 1
\end{array}  \right )
{\bf n}^T(t).
\end{equation}
Equation
(\ref{evolution})
for ${\bf n}^T(t)$ can be replaced by an equivalent
equation
$$
\frac {d}{dt}{\bf u}^T(t)
=\hat{B_u}{\bf u}^T(t),
$$
with ${\bf u}^T(0)={\bf n}^T(0)$
and
$$ \hat{B_u}=\left (
\begin{array}{lcr}
0 & \omega+\omega_1 & 0\\
-(\omega+\omega_1) & 0 &\omega_0\\
0 & -\omega_0 & 0
\end{array}  \right ).
$$
Note that the matrices $\hat{B}_u$ at different times commute because of its
time-independence, from Eq.(\ref{exact-sol}) and (\ref{new-vector}),
the curve ${\bf n}(t)$
is derived exactly as

$$
{\bf n}^T(t)
=\left (
\begin{array}{lcr}
cos\omega t & -sin\omega t & 0\\
sin\omega t & cos\omega t & 0\\
0 & 0 & 1
\end{array}  \right )
$$
\begin{equation}
\label{crown}
\left (
\begin{array}{lcr}
sin^2\chi+cos^2\chi cos\omega_s t
&cos\chi sin\omega_s t
&\frac {1}{2}sin 2\chi(1-cos\omega_s t)
\\
-cos\chi sin\omega_s t 
&cos\omega_s t
& sin\chi sin\omega_s t 
\\
\frac {1}{2}sin 2\chi(1-cos\omega_s t)
&-sin\chi sin\omega_s t
& cos^2\chi+sin^2\chi cos\omega_s t 
\end{array}  \right )
{\bf n}^T(0),
\end{equation}

where $\omega_s=\sqrt{\omega_0^2+(\omega+\omega_1)^2}$
and $\chi=arctg \frac{\omega_0}{\omega_1+\omega}$.
From Eq.(\ref{crown}) and Eq.(\ref{phase}), 
the Pancharatnam phase can be readily
computed analytically or numerically,
which will be useful in studing
the interference effect on the
nonadiabatic noncyclic electronic transport across a mesoscopic
Aharonov-Bohm ring connected to the current leads~\cite{Morpurgo}
(see also, e.g., Sec. IIIA).

For a cyclic evolution, the above result can be further simplified.
The evolution can be cyclic if the frequencies 
$\omega_s$ and $\omega$ 
are commensurable, that is,  
$\omega_s=\frac {m\omega}{k}$ with $m$ and $k$ as irrational integers.
Under this condition,
the corresponding Pancharatnam phase
accumulated in one cycle with  the initial state 
$$          
|\phi(0)\rangle=
\left (
\begin{array}{l}
e^{-i\varphi_i/2}cos(\theta_i/2)\\
e^{i\varphi_i/2}sin(\theta_i/2)
\end{array} 
\right )_\sigma
$$
can be obtained explicitly

\begin{equation}
\label{AA-phase}
\gamma_p(\tau)=\gamma_{AA}(\tau)=m\pi(1-cos\beta)-k\pi
(1-cos\chi cos\beta),
\end{equation}
where $\tau=2k\pi/\omega=2m\pi/\omega_s$,
and $cos\beta=cos\theta_icos\chi+sin\theta_i sin\chi cos\varphi_i$.
If we define an effective magnetic field
in the spherical coordinates
${\bf B}_{eff}(t)=(B_{eff},\chi,\omega t)$
with $B_{eff}=\hbar\omega_s/\mu$,
$\beta$ is just a constant angle between the vectors
${\bf n}(t)$
and ${\bf B}_{eff}(t)$~\cite{Rabi}.
Note that the
evolution given by Eq.(\ref{crown}) is
basically the superposition of two rotations.
The first one is that the effective magnetic field rotates around the $z$
axis with the angle $\chi$ and the angular
frequency $\omega$. The second one is the spin precession around the 
direction of the effective magnetic field with an angle
$\beta$ and a precession
angular frequency $\omega_s$.
The combination of the two rotations leads to a cyclic evolution only if the
frequencies $\omega_s$ and $\omega$ are commensurable.
Obviously, the geometric phases induced by the first
and second rotations are respectively the second and first terms
of the RHS of Eq.(\ref{AA-phase}).
In the adiabatic limit, we have $\chi\rightarrow\alpha$
($\omega/\omega_1\rightarrow 0$), and $\beta\rightarrow 0$
($\varphi_i\rightarrow 0$ and $\theta_i\rightarrow \chi$)
because ${\bf n}(t)$
aligns with ${\bf B}_{eff}(t)$. Therefore,
the adiabatic Berry phase is recovered.

The Pancharatnam phase $\gamma_p(t_f)$ (and $\xi(t_f)$ defined later)
versus
the time $t_f$ is plotted in Fig.1 with
$\varphi_i=\pi/6$, $\theta_i=5\pi/12$,
$\alpha=\pi/3$, $\omega=50\ Hz$,
and $\omega_s=2\omega$ (solid line),
$\sqrt{3}\omega$ (dotted line),
respectively.
If we define a function $\xi(t_f)=\gamma_p(t_f)-\gamma_p(\eta\tau)$
with $\eta=Int[t_f/\tau]$,
we can see from the inset of Fig.1 that
$\xi(t_f)$ is a periodic function of $t_f$ with period $\tau=2\pi/\omega$
for $\omega_s=2\omega$.
Also $\gamma_{AA}=
\gamma_p(t_f)-\gamma_p(t_f-\tau)=0.61$ is the AA phase
for the cyclic evolution. However, the dotted line in Fig.1
does not have the above properties because the evolution is not cyclic
during finite time.

\subsection{Topological transition in a mesoscopic ring subject to an
in-plane magnetic field}

   Recently, Lyanda-Geller investigated the adiabatic Berry phase
induced by the spin-orbit interaction in low dimensional or lowered
symmetry conductors, and proposed an 
interesting phenomenon: topological transition~\cite{Lyanda-Geller}.
Here, we propose that this phenomenon may occur in a mesoscopic ring
subject to an in-plane magnetic field,
which may be easier to be observed.
As an application of Eq.(\ref{phase}), we also analyze whether or not this
topological transition exists in nonadiabatic noncyclic cases.

Consider a mesoscopic ring with radius $r$ connected to
current leads in a static magnetic field, as shown in Fig.2.
We assume that the motion of electrons in the whole system
is ballistic, however,
we include the spin-flip
processes induced by the inhomogeneous magnetic field, which 
is a big merit to consider the Pancharatnam phase rather than 
the cyclic AA phase or the adiabatic Berry phase,
where an artificial restriction
that spin-up and spin-down electrons traverse the ring independently
is required~\cite{Loss,Zhu,Lyanda-Geller}.

   An incoming electron wave incident from the left lead is splitted into
two beams at the left junction and recombined at the right junction
into the outgoing wave through the right lead. As a consequence, the
motion of spin-$\frac{1}{2}$ electron in the textured ring
is equivalent to a quantum spin-$\frac{1}{2}$ in a rotating magnetic field
in time. For a beam of electron wave with Fermi velocity
$V_f=\hbar k_f/m_e$, where $k_f$ is the Fermi wave vector and $m_e$ is the
effective electron mass, the time for electrons to traverse ballistically
one round in the ring is $t_0=\frac {2\pi r}{V_f}$, which is the interval
that the electron moves in the
magnetic field~\cite{Zhu}. In this situation, the Pancharatnam phase mentioned above in
addition to the usual Aharonov-Bohm(AB) phase due to the coupling of electrons to the conventional
electromagnetic gauge potential, is accumulated on the
electron wavefunction. 
In such a system, the quantum transport is significantly affected by
the AB phase and
Pancharatnam phase.
We assume for simplicity the ring to be symmetric. 
Following the method originally
given by B\"{u}ttiker, Imry, and Azbel~\cite{Buttiker},
the transmission coefficient affected by the geometric phase
can be obtained as
\begin{equation}
\label{transmission}
T_g=\frac{2\epsilon^2sin^2(k_f\pi r)(1+cos\gamma)}
{[a^2+b^2cos\gamma-(1-\epsilon)cos(2k_f\pi r)]^2+\epsilon^2sin^2(2k_f\pi r)},
\end{equation}
where $a=\pm(\sqrt{1-2\epsilon}-1)/2$,  $b=\pm(\sqrt{1-2\epsilon}+1)/2$
with $0\leq\epsilon\leq 1/2$,
and $\gamma=2\pi\phi_{AB}/\phi_0+\gamma_p$ 
is the total geometric phase
with $\phi_{AB}$ as 
the AB flux and $\phi_0=h/e$ as the usual flux quantum. 
Here
the parameter $\epsilon$ stands for the coupling strength
of the ring to two leads, and $\epsilon=0$ in the weak coupling
limit while $\epsilon=1/2$ in the strong coupling limit.

The time-dependent Hamiltonian describing the spin motion
in Fig.2 is given by~\cite{Note}
\begin{equation}
\label{TT-Hamiltonian}
\hat{H}=g\mu [\sigma_x(-B_t sin\omega_f t+B_x)+
\sigma_y B_t cos\omega_f t],
\end{equation}
where $\omega_f=2\pi/t_0$ and $g$ is the gyromagnetic ratio. 
A natural basis for $\hat{H}$ consists of $|n_{+}(t)\rangle$
and $|n_{-}(t)\rangle$ that satisfy $\hat{H}(t)|n_j(t)\rangle=
E_j(t)|n_j(t)\rangle$ $(j=+,-)$ is given by
$ \langle n_j(t)|=\frac{1}{\sqrt{2}}( 1 ,
\frac{E_j}{\hbar(\omega_x+i\omega_t exp(i\omega_f t))})$
with corresponding eigenenergies $E_j=j\hbar
\sqrt{\omega_t^2+\omega_x^2-2\omega_t\omega_x sin\omega_f t}$ and
$\omega_{t,x}=g\mu B_{t,x}/\hbar$.
Within the adiabatic approximation,
the Berry phase $\gamma_{Berry}$
accumulated on the wave function is found to be $\gamma_{Berry}=\pi$
for $\omega_x<\omega_t$, $\gamma_{Berry}=\pi/2$ for
$\omega_x=\omega_t$, and $\gamma_{Berry}=0$
for $\omega_x>\omega_t$~\cite{Lyanda-Geller}.
It is interesting to note that the adiabatic Berry phase does not
continuously vary with the magnetic field. Substituting the Berry phase
into Eq.(\ref{transmission}) ($\phi_{AB}=0$),  
the transmission coefficient $T_g$ versus magnetic field can be obtained as
\begin{equation}
\label{TT}
T_g=\left\{ 
\begin{array}{cc}
0, &\ \ for \  \omega_x<\omega_t\\
\frac{8sin^2(k_f \pi r)}{1+8sin^2(k_f \pi r)}, & \ \ for\  \omega_x=\omega_t \\
1, & \ \ for\  \omega_x>\omega_t
\end{array}
\right .
\end{equation}
in the strong coupling limit.
Equation (\ref{TT}) gives a mathematical argument for the existence
of a topological transition in this system which characters
the destructive ($T_g=0$) to
constructive ($T_g=1$) interference in quantum transport affected by
adiabatic Berry phase.
According to the Landauer-B\"{u}ttiker 
formula~\cite{Landauer}, the conductance through the system is
$G=(e^2/\hbar)T_g$. Therefore, the conductance
as a function of either $B_t$ or $B_x$
has steplike character if the other is fixed.
This steplike current-magnetic
field character, which is stemmed from the topological geometric phase,
is referred to as the topological transition. 

Does this topological transition still exist in nonadiabatic noncyclic
cases? To answer this question, we compute
the Pancharatnam phase and substituted it into Eq.(\ref{transmission})
without using the adiabatic approximation or cyclic condition.
For the case that the initial state is an eigenstate,
the transmission coefficient $T_g$ against
$\omega_x/\omega_t$
for $\frac{\omega_t}{\omega_f}=100,\ 10,\ 1$ 
are plotted in Fig.3,
where $\omega_f=10^9 Hz$
\cite{Note}, $k_fr=n+1/2$ with $n$ a non-negative integer.
From Fig.3(a), the rather sharp topological transition occurs 
at $\frac{\omega_x}{\omega_t}=1$ 
for $\frac{\omega_t}{\omega_f}=100$ under which the adiabatic
conditions $\omega_t>>\omega_f$ is well satisfied.
However, for $\frac{\omega_t}{\omega_f}=1,\ 10$(Fig.3(b), (c))
the adiabatic conditions are not well 
satisfied, we can not observe the topological transition.
The above result coincides with a geometric point of view. 
We  can roughly decompose the Pancharatnam
phase into two parts, the phase induced by the magnetic field
trajectory circuit and the spin precession around the magnetic field.
In the adiabatic condition,
the later one is approximately zero because the spin direction
is along the direction of the magnetic field.
Then we only need to analyze the first part.
It was pointed out that the adiabatic Berry phase
for a spin-$\frac{1}{2}$ particle in a magnetic field
is a half of the solid angle
that the magnetic field trajectory subtends at degeneracy
({\sl i.e.}, at $\stackrel{\rightarrow}{B}=0$ point)
~\cite{Berry}. Then $\gamma_{Berry}=0$ for 
$\omega_t<\omega_x$ because the magnetic field trajectory circuit does
not enclose the degeneracy. On the other hand,
$\gamma_{Berry}=\pi$ for $\omega_t>\omega_x$
since the degeneracy is enclosed and the
solid angle of the magnetic circuit is $\pm 2\pi$.
In the nonadiabatic noncyclic cases, however,
the Pancharatnam phase induced by the spin precession
is significant, which oscillates quickly around
$\frac{\omega_x}{\omega_t}=1$,
with the first part almost unchanged.
Therefore, we may 
conclude that
the Pancharatnam phase induced by the spin precession
destroys the topological transition.

Finally, we wish to point out whether or not the topological
transition exists in nonadiabatic noncyclic motion
may be tested by a well designed mesoscopic experiment,
in which $B_t$ may be induced by a long straight current-carrying
wire pass through normally the center of the ring
as shown in Fig.2.
For the ballistic motion in a gold ring with $r\sim 1\mu m$
and $V_f\sim 10^5 m/s$, $g\sim 1$, it is required that the corresponding
field should be $\sim 1\ Tesla$ for $\omega_t\sim \omega_f$ and
$\sim 10^2\ Tesla$ for $\omega_t\sim 100\omega_f$. If the motion in the gold
ring is diffusive, $\omega_f$ is replaced by $\omega_D
=\frac{l}{2\pi r}\omega_f$
($l$ is the elastic mean free path), the required magnitude
field may be less by a factor $\frac{l}{2\pi r}$ (about two orders)
than that predicted for the ballistic case. On the other hand,
$g\sim 15$ in a $GaAs$ ring, the required magnetic field may be less than
$10\ Tesla$ in the case $\omega_t/\omega_f=100$ even for ballistic motion.
Therefore, the results reported here may be tested
in both ballistic and diffusive conditions.

\section{Summary}

A useful formula 
of the noncyclic nonadiabatic geometric phase for a
quantum spin-$\frac {1}{2}$ in an arbitrary 
magnetic field has been formulated exactly,
which can be used in any two-level system.
The formula has been applied to three specific
kinds of magnetic fields.
The evolution equations of the spin-$\frac{1}{2}$ particle  in
an orientated and in a rotating magnetic fields have been
solved respectively,
and the Pancharatnam phases are computed.
We have also found that the nonadiabatic noncyclic phase has
a significant impact on the topological transition
in a mesoscopic system.

\acknowledgements{We gratefully acknowledge
helpful discussions with Prof. Hua-Zhong Li and Dr. Shi-Dong Liang.
This work is supported by a RGC grant of Hong Kong.}

{\bf Figure caption}\newline

 Fig.1 The Pancharatnam phase $\gamma_p(t_f)$
versus the time $t_f$
for $\frac{\omega_s}{\omega}=2$ (solid line),
$\sqrt{3}$ (dotted line). The inset shows
that $\xi(t_f) $ versus the time $t_f$
for $\frac{\omega_s}{\omega}=2$.

Fig.2 A ring connected to
current leads in a uniform external magnetic field $B_x$ and
a tangent magnetic field $B_t$, as described by the
Hamiltonian (\ref{TT-Hamiltonian}).

Fig.3 The transmission coefficients $T_g$ versus the ratio
$\omega_x/\omega_t$ for different
$\omega_t/\omega_f$.

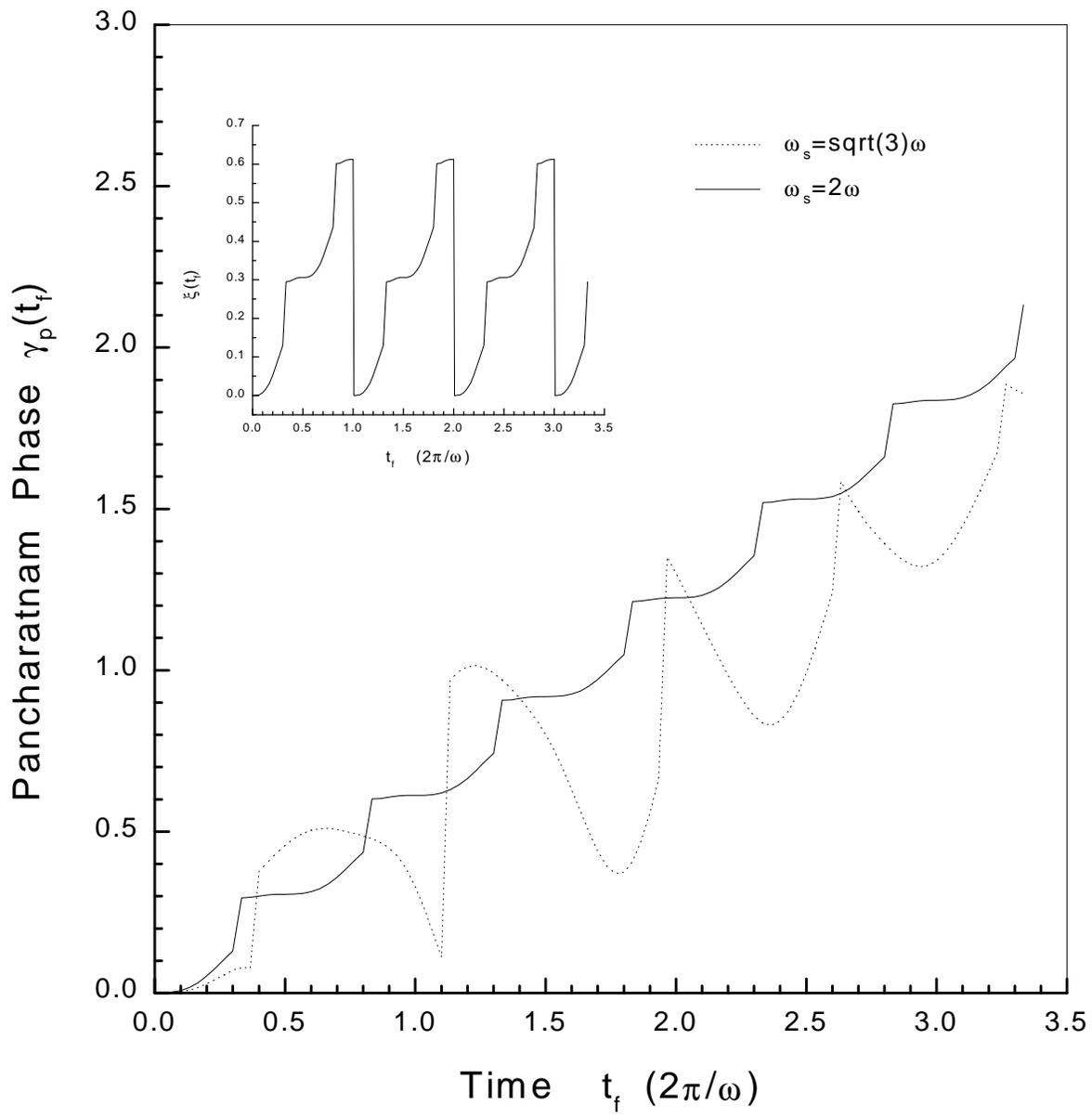
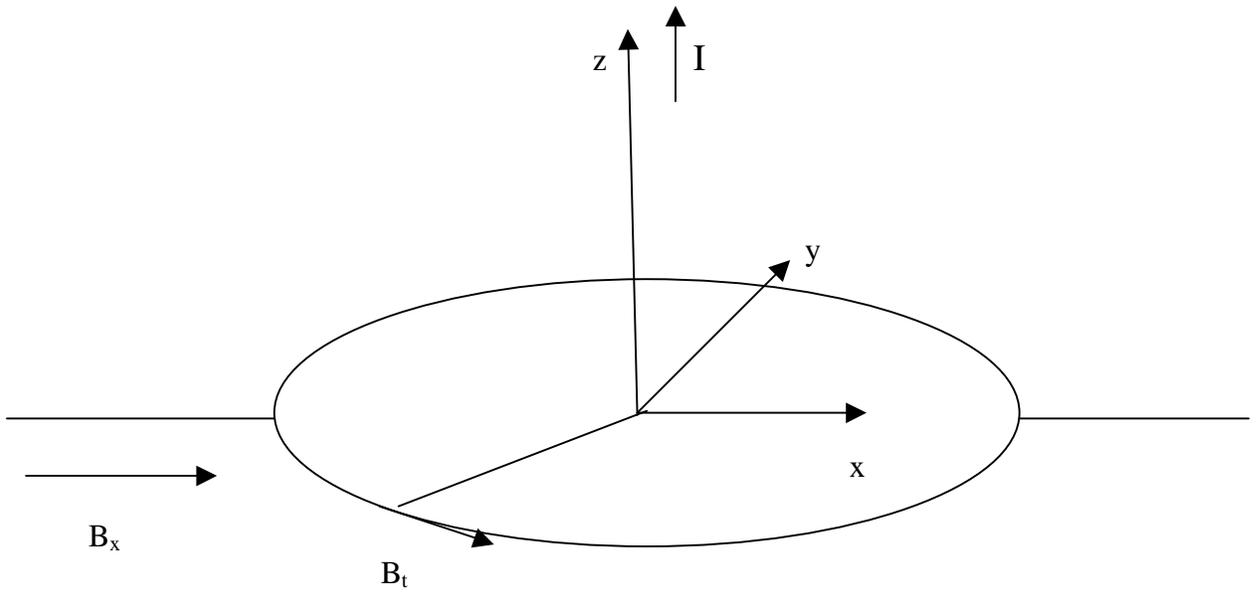
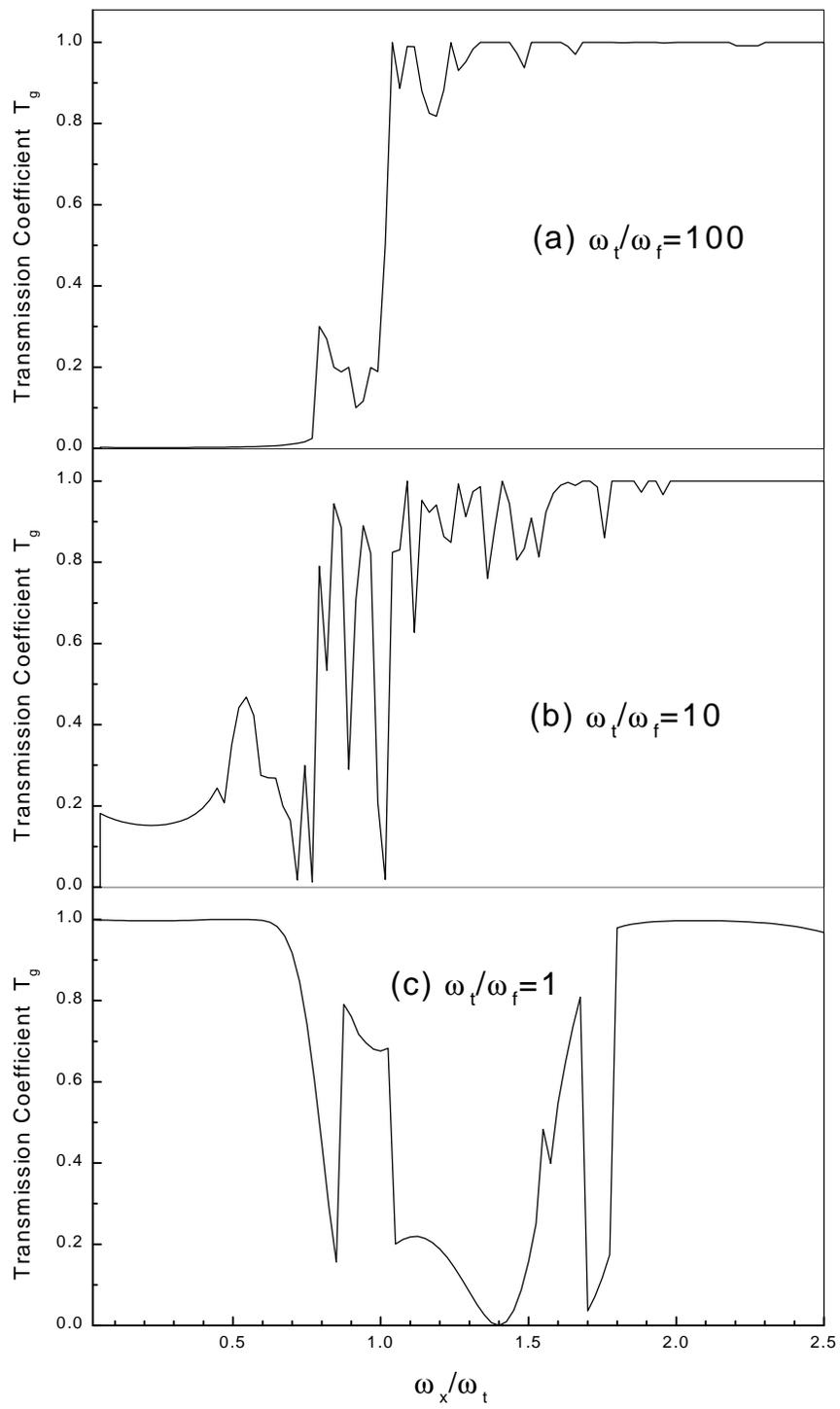
\end{document}